\documentclass[twocolumn,preprintnumbers,nofootinbib,prl,showpacs]{revtex4}

\usepackage{graphicx}
\usepackage{graphics}
\usepackage{dsfont}
\usepackage{epsfig}
\usepackage{amsmath,amssymb,amsthm,amscd}

\usepackage[hypertex]{hyperref}
\newcommand{\beq}{\begin{equation}}
\newcommand{\eeq}{\end{equation}}
\newcommand{\be}{B_\oplus}

\def\be{\begin{equation}}
\def\ee{\end{equation}}
\def\baray{\begin{eqnarray}}
\def\earay{\end{eqnarray}}
\def\ba{\begin{eqnarray}}
\def\ea{\end{eqnarray}}

\begin{document}

\newcommand{\bea}{\begin{eqnarray}}
\newcommand{\eea}{\end{eqnarray}}
\newcommand{\barr}{\begin{array}}
\newcommand{\earr}{\end{array}}

\pagestyle{plain}

\preprint{USTC-ICTS-13-04}

\title{Modular anomaly from holomorphic anomaly \\ in mass deformed $\mathcal{N}=2$ superconformal field theories}

\author{Min-xin Huang
}

\affiliation{Interdisciplinary Center for Theoretical Study,   \\  University of Science and Technology of China, Hefei, Anhui 230026, China
}


\begin{abstract}
We study the instanton partition functions of two well-known superconformal field theories with mass deformations. Two types of anomaly equations, namely, the modular anomaly and holomorphic anomaly, have been discovered in the literature. We provide a clean solution to the long standing puzzle about their precise relation, and obtain some universal formulas. We show that the partition function is invariant under the $SL(2,\mathbb{Z})$ duality which exchanges theories at strong coupling with those of weak coupling. 
\end{abstract}
\pacs{11.15.-q, 11.30.Pb}
\maketitle

Strong coupling effects in quantum field theory are responsible for many interesting physical phenomena, but in general are quite difficult to calculate. The ground breaking works of Seiberg and Witten solve the low energy effective action of $\mathcal{N}=2$ supersymmetric gauge theories in four dimensions, i.e. the prepotential, using its holomorphicity as well as its asymptotic and monodromy properties \cite{SW1, SW2}.  The instanton contributions can also be computed directly by localization technique in the $\Omega$ background and are known as the Nekrasov partition function \cite{Nekrasov:2003}. The $\Omega$ background is a noncommutative deformation of $\mathbb{R}^4$ space by two small parameters $\epsilon_1$, $\epsilon_2$. The leading term in the small $\epsilon$ expansion is exactly the $\mathcal{N}=2$  prepotential, while the higher order terms are the effective couplings of graviphoton fields to the Ricci tensor. 

The effective actions of the $\mathcal{N}=2$ theories can also be computed by topological string theory through the techniques of geometric engineering.  Topological string theory on Calabi-Yau manifolds has been an active area of research and also provides many insights for other areas of physics, such as black hole entropy and supersymmetric gauge theories in four dimensions \cite{mirrorbook}. The A-model topological string partition function depends on Kahler moduli of the Calabi-Yau manifolds and counts holomorphic curves, whose numbers are known as Gromov-Witten invariants.  Mirror symmetry relates the rather difficult A-model problem to the B-model on the mirror Calabi-Yau manifold, which is a deformation theory for the complex structure moduli. The genus-0 B-model prepotential is solved by a Picard-Fuchs linear partial differential equation. The anholomorphic parts of the higher genus amplitudes are determined by the Bershadsky-Cecotti-Ooguri-Vafa (BCOV) holomorphic anomaly equation \cite{BCOV}, and one may also fix the holomorphic ambiguities by appropriate boundary conditions at special points of the moduli space \cite{Huang:2006}. The conventional topological strings correspond to the case of $\epsilon_1+\epsilon_2=0$ in the $\Omega$ background.  The general case inspires the studies of refined topological string theory and the generalized holomorphic anomaly equations \cite{Krefl:2010a, Krefl:2010b, HK2010, Huang:toappear}.

In this paper we consider two well-known superconformal field theories, namely, the $SU(2)$ $\mathcal{N}=2$ gauge theories with an adjoint hypermultiplet and with $N_f=4$ fundamental hypermultiplets. In the first theory the supersymmetry is enhanced to $\mathcal{N}=4$ and the gauge coupling is corrected by neither 
 perturbative nor instanton contributions. For the second theory, the gauge coupling is renormalized by instanton effects, as seen from the Nekrasov partition function. We will turn on mass parameters in the theories, which break the conformal symmetry and keep the $\mathcal{N}=2$ supersymmetry. The first theory with mass deformation is also known as the $\mathcal{N}=2^*$ theory. In both theories the gauge coupling is renormalized by mass deformation. As in \cite{Krefl:2010b, HKK2011}, we shift the mass parameters by $\frac{\epsilon_1+ \epsilon_2}{2}$ in the Nekrasov partition function so that the odd terms in the small $\epsilon$ expansion  vanish.  

We can expand the instanton partition functions of the two theories around the large modulus point in the Coulomb branch, i.e., where the vacuum expectation value of the scalar in the vector multiplet is large. As power series of the flat coordinate $a$,  the coefficients consist of Eisenstein series and Jacobi theta functions as shown in \cite{Minahan:1997, Billo:2011, Billo:2013}. Physically, the quasimodularity comes from the $SL(2,\mathbb{Z})$ duality which acts on the gauge coupling constant. Here the quasimodular forms are weighted homogenous polynomials of the Eisenstein series $E_2, E_4, E_6$. The $E_2$ series transforms with a shift under S-duality so it is not exactly modular. The modular anomaly equations relate the partial derivative of the instanton partition function with respect to $E_2$ to lower order terms.  

On the other hand,  the holomorphic anomaly equation from topological string theory has been used to compute the instanton partition function of $\mathcal{N}=2$ theories \cite{HK, HKK2011}. It was strongly believed that these two approaches are related. However there are apparent differences between them, and no clear derivation from one to the other is available in the literature. In the modular anomaly equation in  \cite{Minahan:1997, Billo:2011, Billo:2013}, the partition functions are expanded around the large Coulomb modulus point and the argument of the quasimodular forms is the bare coupling, while the holomorphic anomaly approach in \cite{HK, HKK2011} gives exact amplitudes at any points of moduli space and the argument of quasimodular forms is the renormalized gauge coupling. Furthermore, the modular anomaly appears already at genus 0 while the holomorphic anomaly appears only at higher genus.  In this paper we shall fill in the gap and derive the equivalence of the anomaly equations. Along the way, we also obtain some nice formulas which will be useful elsewhere \cite{Huang:toappear}. 

A similar issue also appears in the studies of topological strings on a class of elliptically fibered Calabi-Yau manifolds \cite{Alim:2012, Klemm:2012}. When a $\mathbb{P}^1$ is blown up in the base of these Calabi-Yau models, one can perform T-duality on the fiber and geometrically engineer $\mathcal{N}=4$ topological string theory on the local half K3 manifold \cite{VW}. Here the modular anomaly equation first appeared in \cite{Minahan:1997a} for genus-0 case, and has been generalized in, e.g., \cite{MNVW, HST}. It is also strongly believed that the modular anomaly comes from the BCOV holomorphic anomaly equation \cite{BCOV}, and an argument using the BCOV relations for higher point functions is presented in \cite{Klemm:2012}. In a related paper \cite{Huang:toappear} we will resolve this long standing issue.  

First we consider the case of $\mathcal{N}=2^*$ theory. The Seiberg-Witten curve is 
\begin{eqnarray} \label{SWcurve}
y^2 = 4x(x+u+\frac{m^2}{4}) (x+\tilde{q}u+\frac{\tilde{q}^2m^2}{4}),
\end{eqnarray}
where $m$ is the mass of the adjoint hypermultiplet, $u$ is the Coulomb modulus parameter related to the expectation value of the  scalar in the $\mathcal{N}=2$ vector multiplet, and the parameter $\tilde{q} := \frac{\theta_2(\tau_0)^4}{\theta_3(\tau_0)^4}$ is related to the bare gauge coupling constant $\tau_0$ by Jacobi theta functions. 

We can shift the $x$ parameter to transform the curve into the Weierstrass form $y^2= 4x^3 -g_2 x-g_3$, where the coefficients $g_2, g_3$ are some polynomials of $u, \tilde{q}, m$. The renormalized gauge coupling $\tau$ is the elliptic parameter of the curve \cite{Zagier2008}, and can be determined by the $J$-function 
\begin{eqnarray} \label{effective}
J(\tau)= \frac{E_4(\tau)^3}{E_4(\tau)^3-E_6(\tau)^2} = \frac{g_2 ^3}{ g_2^3-27g_3^2}. 
\end{eqnarray}
The period  $a$ of the Seiberg-Witten curve  is calculated by  \cite{BS}
\begin{eqnarray} \label{dau}
 \frac{da}{du} =\sqrt{-\frac{1}{18} \frac{g_2 }{g_3 } \frac{E_6(\tau)}{E_4(\tau)}}. 
 \end{eqnarray}
It is easy to see that in the large modulus limit $u\rightarrow \infty$, or equivalently massless limit $m=0$, the renormalized gauge coupling $\tau$ is the same as the bare coupling $\tau_0$. However, in general they are different.  

We can expand the Nekrasov partition function $Z(\epsilon_1, \epsilon_2,a, \tau_0)$ for small $\epsilon$ as
\begin{eqnarray}
 \log (Z) =\sum_{n,g=0}^{\infty} (\epsilon_1+\epsilon_2)^{2n} (\epsilon_1\epsilon_2)^{g-1} F^{(n,g)} (a,\tau_0),
\end{eqnarray}
where in our notation, $F^{(n,g)} (a,\tau_0)$ includes the perturbative and instanton contributions. There is also an additional classical contribution at the leading term $F^{(0,0)}_{classical} = - 2\pi i \tau_0 a^2$. We can further expand the amplitudes $F^{(n,g)} (a,\tau_0)$ around the large Coulomb modulus point where $a\sim \infty$ as 
\begin{eqnarray}  \label{expandF}
F^{(n,g)} (a,\tau_0) =f^{(n,g)}_0(\tau_0) \log(a) + \sum_{l=1}^{\infty} \frac{f^{(n,g)}_l(\tau_0)}{a^{2l} },
\end{eqnarray}   
where the coefficients of the logarithmic terms  are $f^{(0,0)}_0= - m^2$, $f^{(1,0)}_0=  \frac{1}{4}$ and vanish for other cases. It was found in \cite{Minahan:1997, Billo:2013} that $f^{(n,g)}_l(\tau_0)$ are quasimodular forms of $\tau_0$ of weight $2l$, and satisfy the modular anomaly equation 
\begin{eqnarray} \label{modular anomaly}
&& \partial_{E_2(\tau_0)} F^{(n,g)} (a,\tau_0) 
= \frac{1}{48} [\partial_a^2 F^{(n,g-1)} \\ \nonumber  && + \sum_{n_1=0}^n\sum_{g_1=0}^g 
(\partial_a F^{(n_1,g_1)} ) (\partial_a F^{(n-n_1,g-g_1)} ) ],
\end{eqnarray}
where the first term on the rhs is absent for $g=0$.  

The prepotential $F^{(0,0)}$ without the  classical contributions  is determined by the effective coupling  $\partial_a^2 F^{(0,0)} = -4\pi i (\tau-\tau_0) $.  Using (\ref{dau}) we find the relation for the dual period $a_D=\partial_a F^{(0,0)}$,  
\begin{eqnarray} \label{dual period}
 \frac{d a_D }{du} =  -4\pi i (\tau-\tau_0) \sqrt{-\frac{1}{18} \frac{g_2 }{g_3 } \frac{E_6(\tau)}{E_4(\tau)}}. 
 \end{eqnarray}

The holomorphic anomaly approach is formulated with the effective gauge coupling $\tau$ in (\ref{effective}). The higher genus amplitudes  $F^{(n,g)}$ for $n+g\geq 2$ are polynomials of $X=\frac{E_2(\tau) E_4(\tau)}{E_6(\tau)}$ with a rational function of $u, \tilde{q}$ as coefficients.  The holomorphic anomaly equation is 
\begin{eqnarray} \label{holomorphic anomaly}
&& \frac{ E_4(\tau)}{E_6(\tau)} \partial_{X } F^{(n,g)} (X,u, \tilde{q} ) 
= \frac{1}{48} [\partial_a^2 F^{(n,g-1)}  \\ \nonumber  && + (\sum_{n_1=0}^n\sum_{g_1=0}^g )^{\prime} 
(\partial_a F^{(n_1,g_1)} ) (\partial_a F^{(n-n_1,g-g_1)} ) ],
\end{eqnarray}
where the prime denotes  the exclusion of the cases $(n_1,g_1)=(0,0)$, $(n_1,g_1)=(n,g)$ in the sum. The two equations (\ref{modular anomaly}, \ref{holomorphic anomaly}) are similar but not quite the same. We shall show that they are indeed equivalent. 

Suppose $P_k$ is a rational function of Eisenstein series and theta functions with weight $k$; then there is a useful formula on the commutation relation of $E_2$ and $\tau$ derivatives, 
\begin{eqnarray} \label{formula}
\frac{1}{2\pi i } (\partial_{E_2}\partial_\tau^n -\partial_\tau^n \partial_{E_2} )P_k  = \frac{n(k+n-1)}{12} \partial_\tau^{n-1} P_k ,
\end{eqnarray} 
which follows from the Ramanujan identities for Eisenstein series and theta functions. For more details see e.g. the review \cite{Zagier2008}. 

We can find the relation between $E_2(\tau)$ and $E_2(\tau_0)$ derivatives. From the relation (\ref{effective}) we can expand the rhs for large $u$ and compute the effective coupling as a series expansion with quasimodular forms and theta functions of bare coupling as coefficients:  
\begin{eqnarray} \label{expansion}
2\pi i (\tau -\tau_0) &=&  -\frac{m^2}{2u \theta_3(\tau_0)^4} -\frac{m^4}{48 u^2 \theta_3(\tau_0)^8} [E_2(\tau_0) 
 \nonumber \\ &&  - 4\theta_2(\tau_0)^4 -4\theta_3(\tau_0)^4 ]
+ \mathcal{O} (\frac{1}{u^3}) . 
\end{eqnarray}
Using the formula (\ref{formula}) and the Taylor expansion for a rational function of quasimodular forms of $\tau$, we find 
\begin{eqnarray} \label{formula10}
&& ~\partial_{E_2(\tau_0)} P_k(\tau)  = \partial_{E_2(\tau)} P_k(\tau) +\frac{k}{12}2\pi i (\tau -\tau_0) P_k  \nonumber \\
&& + ( \partial_{\tau} P_k(\tau)) [ \partial_{E_2(\tau_0)}(\tau-\tau_0) +\frac{2\pi i}{12} (\tau-\tau_0)^2 ]. 
\end{eqnarray}
We  note that the $E_2$ derivatives of the Jacobi theta functions vanish. Applying the above formula to (\ref{effective}) which has zero modular weight, we find the simple formula 
\begin{eqnarray} \label{simple formula}
\partial_{E_2(\tau_0)}(\tau-\tau_0) = - \frac{2\pi i}{12} (\tau-\tau_0)^2, 
\end{eqnarray}
which was also derived in \cite{Minahan:1997} and can also be checked explicitly using the expansion (\ref{expansion}). So the formula (\ref{formula10}) simplifies to 
\begin{eqnarray} \label{formula12}
\partial_{E_2(\tau_0)} P_k(\tau)  = \partial_{E_2(\tau)} P_k(\tau) +\frac{k}{12}2\pi i (\tau -\tau_0) P_k 
\end{eqnarray}

For weight-0 quasimodular forms, the $E_2(\tau_0)$ and $E_2(\tau)$ derivatives are the same. For example, for  $X=\frac{E_2(\tau) E_4(\tau)}{E_6(\tau)}$  we find $\partial_{E_2(\tau_0)} X(u, \tau_0)  = \frac{ E_4(\tau)}{E_6(\tau)}$.
 
We can compute the $E_2(\tau_0)$ derivative of the periods $a, a_D$   as a function of $u$ and $\tau_0$. Applying  (\ref{formula12}) to (\ref{dau}) we find 
\begin{eqnarray}
\partial_u \partial_{E_2(\tau_0)} a(u, \tau_0) = \frac{2\pi i (\tau-\tau_0)}{12}  \sqrt{-\frac{1}{18} \frac{g_2 }{g_3 } \frac{E_6(\tau)}{E_4(\tau)}}. 
\end{eqnarray} 
Comparing with (\ref{dual period}), up to an integration constant of $u$, which can be easily checked to be zero, we find 
\begin{eqnarray} \label{aE2}
 \partial_{E_2(\tau_0)} a(u, \tau_0) = -\frac{1}{24} \partial_a F^{(0,0)} . 
\end{eqnarray} 
For the dual period (\ref{dual period}), we use the formulas (\ref{simple formula}) and (\ref{formula12}) and find a vanishing result after a nice cancellation: 
\begin{eqnarray}  \label{aDE2}
 \partial_{E_2(\tau_0)} a_D (u, \tau_0) = 0
\end{eqnarray} 

The modular anomaly equation (\ref{modular anomaly}) for genus-0 case \cite{Minahan:1997} can be deduced from the results (\ref{aE2}, \ref{aDE2}). This follows by taking the derivative of $a$ on both sides of the equation (\ref{modular anomaly})  and using the chain rule for taking the derivative. 

For the higher genus, we can again use the chain rule for taking the derivative and (\ref{aE2}) to move the genus 0 and genus $(n,g)$  amplitudes from the rhs to the lhs. So the modular anomaly equation (\ref{modular anomaly})  is equivalent to 
\begin{eqnarray} \label{equivalent}
&& \partial_{E_2(\tau_0)} F^{(n,g)} (u,\tau_0) 
= \frac{1}{48} [\partial_a^2 F^{(n,g-1)}   \\ \nonumber && + (\sum_{n_1=0}^n\sum_{g_1=0}^g )^{\prime} 
(\partial_a F^{(n_1,g_1)} ) (\partial_a F^{(n-n_1,g-g_1)} ) ],
\end{eqnarray}
where on the lhs we are now taking the $E_2(\tau_0)$ derivative with $u$ fixed, and the prime in the sum excludes the cases $(n_1,g_1)=(0,0)$, $(n_1,g_1)=(n,g)$ as in the holomorphic anomaly equation (\ref{holomorphic anomaly}).  Using the fact that $\partial_{E_2(\tau_0)} F^{(n,g)} (u, \tau_0) = \frac{ E_4(\tau)}{E_6(\tau)} \partial_X F^{(n,g)} (X,u, \tilde{q} )$, we derive the equivalence of anomaly equations (\ref{modular anomaly}) and  (\ref{holomorphic anomaly}). 

The case of genus 1 requires some special care. The exact formulas from the holomorphic anomaly  are 
\begin{eqnarray} \label{genus one}
F^{(1,0)} &=& \frac{1}{48} \log(\Delta), \nonumber \\ 
F^{(0,1)} &=& -\frac{1}{2} \log(\frac{da }{du}) -\frac{1}{24} \log(\Delta) ,
\end{eqnarray}
where $\Delta= g_2^3-27 g_3^2$ is the discriminant, which in this case is a perfect square of a cubic polynomial of $u$. We can verify the modular anomaly equation (\ref{equivalent}). The case of $ F^{(1,0)}$ is obvious. For the case of  $F^{(0,1)}$, we use the formulas (\ref{dau}) and (\ref{formula12}), with a nonzero modular weight:
\begin{eqnarray}
\partial_{E_2(\tau_0)} F^{(1,0)} (u,\tau_0)  &=& 0 ,  \\ \nonumber 
\partial_{E_2(\tau_0)} F^{(0,1)} (u,\tau_0) &=&  -\frac{2\pi i}{24} (\tau-\tau_0) =\frac{1}{48} \partial_a^2 F^{(0,0)} . 
\end{eqnarray}

We shall also show that the coefficients $f^{(n,g)}_l(\tau_0)$ in the expansion (\ref{expandF}) are quasimodular forms, i.e., polynomials of Eisenstein series, though naively they could contain Jacobi theta functions. Under the $SL(2,\mathbb{Z})$ action on the bare coupling $\tau_0$, the $\tilde{q}$ parameter transforms according to the 
well-known rule for theta functions. We also perform simultaneous transformations for the Coulomb modulus:  
\begin{eqnarray} \label{theta transform}
&&\textrm{T-duality}: ~~~ \tilde{q}\rightarrow -\frac{\tilde{q}} {1-\tilde{q}}, ~~ u\rightarrow \frac{1}{1-\tilde{q}} (u+ \tilde{q}  \frac{m^2}{2}),  \nonumber \\
&&\textrm{S-duality}: ~~~ \tilde{q}\rightarrow 1-\tilde{q}, ~~ u\rightarrow  -u - \frac{m^2}{2} . 
\end{eqnarray} 

The Seiberg-Witten curve (\ref{SWcurve}) remains the same under the transformations (\ref{theta transform}) by the following shifts and scaling for parameters $x,y$ of the curve: 
\begin{eqnarray}
&&\textrm{T-duality}: ~~~ x\rightarrow \frac{x+\tilde{q}u+\frac{\tilde{q}^2 m^2}{4}}{(1-\tilde{q})^2},~~ y\rightarrow \frac{y}{(1-\tilde{q})^6} , \nonumber  \\
&&\textrm{S-duality}: ~~~ x\rightarrow x+u+\frac{m^2}{4},~~ y\rightarrow y . 
\end{eqnarray}
So these transformations (\ref{theta transform})  scale $g_2\rightarrow (1-\tilde{q})^{-4} g_2$,    
 $g_3\rightarrow (1-\tilde{q})^{-6} g_3$ for T-duality, and keep $g_2$, $g_3$ invariant for S-duality, as can also be explicitly checked with the formulas for $g_2, g_3$. In both cases the $J$-function on the rhs in (\ref{effective}) is invariant. We can then identity the $SL(2,\mathbb{Z})$ transformations of the effective coupling $\tau$ with those of the bare coupling $\tau_0$, together with the above transformation rules  for $u$ in (\ref{theta transform}). 

We can find the transformation rules for period $a$ using (\ref{dau}). For T-duality, the transformation of the rhs cancels that of the lhs from $du$, so $a$ is invariant. For S-duality, there is a factor of $\tau$ due to the nonzero modular weight, so  $ \frac{d a}{du} \rightarrow \tau \frac{d a}{du} $. We see that $a^2$ transforms basically with weight $2$, but with the effective coupling $\tau$ factor instead of the bare coupling $\tau_0$. We can take a limit $\tau_0\rightarrow \infty$  and keep the Eisenstein series and theta functions fixed. In this limit we have $\tau\sim \tau_0$ from the expansion (\ref{expansion}), and the shift in the $E_2(\tau_0)$ S-duality transformation vanishes. The coefficients $f^{(n,g)}_l(\tau_0)$ have no pole in the $\tau_0$ plane. If $F^{(n,g)}$ is invariant under the $SL(2,\mathbb{Z})$ transformations (\ref{theta transform}) for $\tilde{q}$ and $u$, then the theta functions in the $E_2$ independent parts in $f^{(n,g)}_l(\tau_0)$ must combine into modular forms, i.e. polynomials of $E_4(\tau_0)$ and $E_6(\tau_0)$. 

The modular invariance of $F^{(n,g)}$ in the limit $\tau_0\sim \infty$ can be straightforwardly checked using the formulas in \cite{HKK2011}. For the genus-0 case, we have $\partial_a^2 F^{(0,0)} =-4\pi i (\tau-\tau_0) $, which has modular weight $-2$, so $F^{(0,0)}$ is modular invariant. For the genus-1 case, we can also easily check the modular invariance (up to some constants independent of $a$) using the formulas (\ref{genus one}) and (\ref{dau}). For higher genus cases $n+g\geq 2$, the $X=\frac{E_2(\tau) E_4(\tau)}{E_6(\tau)}$ is modular invariant in the limit $\tau_0\sim \infty$; we apply the modular transformation rules (\ref{theta transform})  for $\tilde{q}$ and $u$ to the polynomial formulas for  $F^{(n,g)}$ and check the modular invariance explicitly up to genus 3. Actually, the modular invariance is somewhat expected since the formalism in \cite{HKK2011} depends mostly on the $J$-function and the discriminant, which are invariant under the modular transformations.

The analysis of the $N_f=4$ theory is similar. In this case, the Seiberg-Witten curve is more complicated as it contains four mass parameters. The curve can be found, e.g., in \cite{HKK2011}. Here it is the bare coupling $q_0=\exp(2\pi i \tau_0)$ instead of its theta functions that appears in the Seiberg-Witten curve. We should define an auxiliary parameter $\tilde{q}$ such that $q_0=\frac{\theta_2(\tilde{q})^4}{\theta_3(\tilde{q})^4}$. The roles of the parameters $\tilde{q}$ and $q_0$ are now exchanged comparing with the previous case of the $\mathcal{N}=2^*$ theory. We again find the $\tau$ of the curve goes like $\tau\sim \tilde{\tau}$ in the large Coulomb modulus or massless limit. Since $q_0\sim \tilde{q}^{\frac{1}{2}}$ for small $\tilde{q}$, the $\tau$ of the curve is actually twice the renormalized gauge coupling. Except for the factor of $2$, the formulas can be made the same as in the $\mathcal{N}=2^*$ theory by simply switching the parameters $q_0$ and $\tilde{q}$, and we will not repeat the details again . For example, in the anomaly equations (\ref{modular anomaly}) and (\ref{holomorphic anomaly}), the factor of $48$ should become $24$ for the $N_f=4$ theory, and the derivative in the lhs of (\ref{modular anomaly})  is with respect to $E_2(\tilde{\tau})$. 

There is an additional subtlety in the $N_f=4$ case. Here the coefficients $f^{(n,g)}_l(\tau_0)$ in the expansion (\ref{expandF}) contain Jacobi theta functions which cannot be combined into modular forms,  as can be seen explicitly in the formulas in \cite{Billo:2011, Billo:2013}.   
We should also transform the mass parameters in addition to the Coulomb modulus $u$ in order to keep the Seiberg-Witten curve invariant. It is straightforward to find the transformations analog to (\ref{theta transform}), and explain the quasimodularity and the patterns of theta functions in $f^{(n,g)}_l(\tau_0)$.

\noindent \textit{Acknowledgements}: We thank Albrecht Klemm for careful readings of the draft and collaborations on related papers. MH is supported by the ``Young Thousand People" plan by the Central Organization Department in China.

\end{document}